\documentclass[twocolumn,showpacs,preprintnumbers,amsmath,amssymb,pre,aps,10pt]{revtex4-2}

\usepackage{graphicx}
\usepackage{dcolumn}
\usepackage{bm}
\usepackage{mathrsfs}
\usepackage{color}
\usepackage[table]{xcolor}

\usepackage{scalerel}
\usepackage{tikz}
\usetikzlibrary{svg.path}

\definecolor{orcidlogocol}{HTML}{A6CE39}
\tikzset{
  orcidlogo/.pic={
    \fill[orcidlogocol] svg{M256,128c0,70.7-57.3,128-128,128C57.3,256,0,198.7,0,128C0,57.3,57.3,0,128,0C198.7,0,256,57.3,256,128z};
    \fill[white] svg{M86.3,186.2H70.9V79.1h15.4v48.4V186.2z}
                 svg{M108.9,79.1h41.6c39.6,0,57,28.3,57,53.6c0,27.5-21.5,53.6-56.8,53.6h-41.8V79.1z M124.3,172.4h24.5c34.9,0,42.9-26.5,42.9-39.7c0-21.5-13.7-39.7-43.7-39.7h-23.7V172.4z}
                 svg{M88.7,56.8c0,5.5-4.5,10.1-10.1,10.1c-5.6,0-10.1-4.6-10.1-10.1c0-5.6,4.5-10.1,10.1-10.1C84.2,46.7,88.7,51.3,88.7,56.8z};
  }
}

\newcommand\orcidicon[1]{\href{https://orcid.org/#1}{\mbox{\scalerel*{
\begin{tikzpicture}[yscale=-1,transform shape]
\pic{orcidlogo};
\end{tikzpicture}
}{|}}}}

\usepackage{hyperref} 

\begin{document}

\title{Confined modes of single-particle trajectories induced by stochastic resetting}

\author{Aleksander A. Stanislavsky\orcidicon{0000-0003-4420-047X}$^{1}$}  
\email{a.a.stanislavsky@rian.kharkov.ua}
\author{Aleksander Weron\orcidicon{0000-0002-8584-9876}$^2$}    
\email{aleksander.weron@pwr.edu.pl}
\affiliation{$^1$Institute of Radio Astronomy, 4 Mystetstv St., 61002 Kharkiv, Ukraine\\
$^2$Faculty of Pure and Applied Mathematics, Hugo Steinhaus Center, Wroc{\l}aw University of Science and Technology, Wyb.
Wyspia\'{n}skiego 27, 50-370 Wroc{\l}aw, Poland}
\date{\today}

\begin{abstract}
Random trajectories of single particles in living cells contain information about the interaction between particles, as well as, with the cellular environment. However, precise consideration of the underlying stochastic properties, beyond normal diffusion, remains a challenge as applied to each particle trajectory separately. In this paper, we show how positions of confined particles in living cells can obey not only the Laplace distribution, but the Linnik one. This feature is detected in experimental data for the motion of G proteins and coupled receptors in cells, and its origin is explained in terms of stochastic resetting. This resetting process generates power-law waiting times, giving rise to the Linnik statistics in confined motion, and also includes exponentially distributed times as a limit case leading to the Laplace one. The stochastic process, which is affected by the resetting, can be Brownian motion commonly found in cells. Other possible models producing similar effects are discussed.
\end{abstract}

                             
\maketitle

\section{Introduction}
Combination of the live-cell single-molecule imaging with single-particle tracking (SPT) methods has allowed a revolutionary breakthrough for the quantitative analysis of dynamic processes in living cells~\cite{mgp15,sjbwlc17,lgkc21,mgvgm21}. This approach gives the visualization of the movement of  individual ``particles'', the latter being single molecules,  macromolecular complexes, as well as, viruses or organelles in physiological conditions~\cite{csc14}. As such, it has been crucial to study the mechanisms of intracellular transport, cell membrane dynamics, and viral infection~\cite{sj97,w09,bz07}.  Stochastic processes associated with the movement of particles are directly affected by interactions that occur with other cellular structures or components \cite{msm97}.  Therefore, single-particle dynamics often deviates from Brownian motion and exhibits heterogeneous behavior characterized by changes in diffusion, transient confinement, immobilization or anomalous diffusion~\cite{bkv18,wjbsc18,jklosw20}. The development of theoretical frameworks for the robust analysis of random trajectories implemented in biological scenarios is thus of fundamental importance to understand molecular mechanisms of interaction \cite{bkjbgw12}.  

As an example, detection of the transient confinement with high precision requires the knowledge of position or displacement statistics~\cite{sw20a}. Restrained trajectories have shown to obey the Gaussian statistics, as well as, the Laplace one (see, for example, \cite{ewowals12,sw21a}). The Gaussian confinement is often described by the well-known Ornstein-Uhlenbeck (OU) model~\cite{ou30}. In this case, the stationary probability distribution function (PDF) is Gaussian, and the random trajectories follow Gaussian statistics with an obvious physical interpretation. Unfortunately, the OU model, based on the ordinary Langevin equation with a harmonic potential, does not provide a description for confined trajectories with the Laplace statistics.

An alternative scenario, leading to confinement of single-particle trajectories with the Laplace statistics, can be based on the stochastic resetting methodology. The resetting of a stochastic process describes evolution of a stochastic system that is returned repeatedly to a steady (or equilibrium) state~\cite{em11}, as it occurs in target search with home returns~\cite{ems20,pkr20,sw22}. If the resetting process is Poissonian and independent of the random motion that undergoes resetting, then the latter has a steady PDF. For trajectories undergoing Brownian motion with Poissonian resetting, the steady PDF has the Laplace form. It is noteworthy that subordinated Brownian processes, leading to subdiffusion, under Poissonian resetting also produces a stationary state with the Laplace distribution, but with another scale parameter~\cite{sw21}. In contrast, the engagement of L\'evy motion with the stochastic resetting produces confinement with the Linnik statistics, in which the Laplace one is a particular case (see \cite{kkp01}). Experimental works have provided examples of the occurrence of confinement with Laplace statistics~\cite{ewowals12,sw20a,sw21a}, but to our knowledge no examples of confined trajectories with Linnik PDFs have been reported.  

In this article, considering the frequent occurrence of stochastic resetting in biological systems, we propose an analysis pipeline to robustly determine the testing and apply it to several experimental datasets. Our study shows that the Linnik statistics does occur in single-particle trajectories of both G proteins and coupled receptors in living cells. After providing a brief mathematical background of the stochastic resetting, we describe the analytical framework and its application to the experiments that led to the quantification of confined trajectories and their segments following the Laplace and Linnik statistics. We discuss the results and further propose a possible explanation for their occurrence. In our consideration, the transient behavior means that the set of trajectories contains segments in different diffusive modes: free, confined, and others. Each segment obeys one of the modes, however trajectories not segmented can be also found. The case when either segmental or non-segmental trajectories have different statistics indicates the inhomogeneity of the medium and the mode contribution ratio determines how heterogeneous the medium is. If all the trajectories follow the Brownian motion, the medium is homogeneous.

\section{Models of confined modes}
There are at least three ways to get a stationary PDF for stochastic processes, namely
\begin{itemize}
\item subordination of random processes;
\item stochastic differential equations (SDEs) with an attractive potential;
\item stochastic resetting.
\end{itemize}
Each of them may be considered as a mode leading to confinement. Sometimes, they have similarities in limit cases, but, generally, the  limits are different.  Let us consider the above approaches and their features below.

\subsection{Model 1: subordination of random processes} 
The subordination consists of time randomization of a stochastic process with the help of another independent random process~\cite{b49}. Confined trajectories obeying the Laplace PDF can be produced by the use of a specific subordinator, closely related to the one providing tempered subdiffusion. The conjugate property of Bernstein functions connects the tempered subdiffusion with the confinement \cite{sw21}. Interpretation of anomalous diffusion tending to the confinement is that diffusive motion, accompanied by multiple-trapping events with infinite mean sojourn time, is transformed into pure jumps, restricted in a confined environment. If the Laplace exponent of a tempered $\alpha$-stable process is $(s+\chi)^{\alpha}-\chi^{\alpha}$, where $\chi$ is a positive constant and $0 < \alpha < 1$, then its conjugate partner from the set of Bernstein functions takes the form $s/((s+\chi)^{\alpha}-\chi^{\alpha})$ \cite{sw20}. The PDF of the operational time is easy to present in the Laplace transform with respect to $t$, i.e.,
\begin{equation}
\tilde{f}(\tau,s)=\frac{1}{(s+\chi)^{\alpha}-\chi^{\alpha}}\,e^{-\tau s/((s+\chi)^{\alpha}-\chi^{\alpha})}\,.\label{eq1}
\end{equation}
The propagator of such a subordinated process can be written in the integral form
\begin{equation}
p(x,t) = \int^\infty_0 h(x,\tau)\,f(\tau,t)\,d\tau\,,\label{eq2}
\end{equation}
where $h(x,\tau)$ is the PDF of a parent process, whereas $f(\tau,t)$ is the PDF of a directing one. If the ordinary Brownian motion is a parent process, and the directing process is described by Eq. (\ref{eq1}), it is not difficult to find a stationary distribution in the Laplace form explicitly
\begin{equation}
p_B(x,\infty)=\frac{1}{\sqrt{2D\alpha\chi^{\alpha-1}}}\,\exp\left(-\frac{2|x|}{\sqrt{2D\alpha\chi^{\alpha-1}}}\right)\,,\label{eq3}
\end{equation}
where $D$ is the diffusivity constant for the Brownian motion. Any subordinated non-Brownian motion, in which the subordinator is defined by the Laplace exponent conjugate to a tempered $\alpha$-stable process, has a confined probability distribution itself \cite{sw20a,sw21a}. Their forms are simpler to present through the Fourier transform, giving characteristic functions. This procedure covers a wide class of geometrically infinitely divisible distributions as a confined case of the non-Brownian motion subordinated by a special subordinator responsible for the confinement. 

\subsection{Model 2: coupled SDEs with a potential}\label{sdes}
The coupled Langevin equations for position $x_t$ and diffusivity $D_t$ opens new possibilities in description of the confinement. The system of SDEs reads
\begin{equation}
\left\{
    \begin{array}{ll}
dx_t=-\frac{1}{\tau_{x}}(x_t-\bar{x})dt+\sqrt{D_t}d W_t^{(1)} \\
dD_t=-\frac{1}{\tau_{D}}(D_t-\bar{D})dt+\sigma\sqrt{D_t}dW_t^{(2)}
\end{array}\right.\,,\label{eq4}
\end{equation}
where $\tau_{x}$ and $\tau_{D}$ are the correlation times for $x$ and $D$, $\bar{x}$ and $\bar{D}$ denote the average position and diffusivity, whereas $\sigma$ is the ``speed'' of fluctuations for the diffusion coefficient \cite{lscw22}. Here, $W_t^{(1)}$ and $W_t^{(2)}$ are independent Wiener processes. Note that the second equation of the system (\ref{eq4}) is independent of the first, whereas the latter is ``driven'' by the former. Moreover, the former corresponds to a Cox-Ingersoll-Ross (CIR) process \cite{cir85}. The stationary solution of the CIR SDE is presented, for example, in \cite{brm2020}. This PDF is expressed in terms of the gamma distribution. Next, the stationary state for the first SDE in the system (\ref{eq4}) leads to the Gaussian PDF, if its standard deviation became constant as in the ordinary OU model. To find the stationary PDF $F_\lambda(x)$ of the coupled equations (\ref{eq4}), we integrate the Gaussian and the gamma PDFs over $D$. Consequently, the stationary PDF takes the following form:
\begin{eqnarray}
F_\lambda(x) &=& \frac{2\eta^{-\lambda-1/2}}{\Gamma(\lambda)\sqrt{\pi}}|x-\bar{x}|^{\lambda-1/2}\nonumber\\
&\times&K_{\lambda-1/2}\left(\frac{2|x-\bar{x}|}{\eta}\right)\,,\label{eq5}
\end{eqnarray}
where $K_\nu(z)$ is the modified Bessel function \cite{as65} (here $\nu = \lambda-1/2$), $\eta = \sqrt{\tau_x\tau_D\sigma^2}$, and $\lambda=\frac{2\bar D}{\tau_D\sigma^2}$. The index $\lambda-1/2$ has the meaning of the shape parameter \cite{k82}. It is also interesting to mention the characteristic function for this PDF. It reads
\begin{equation}
\int_{-\infty}^\infty F_\lambda(x)\,\cos(qx)\,dx = \frac{e^{iq\bar x}}{\left(1+q^2\eta^2/4\right)^\lambda}\,,\label{eq6}
\end{equation}
showing similarity with both the Laplace and the Linnik distributions. Note, if $\lambda =1$, then Eq.(\ref{eq5}) takes the form of the Laplace PDF
\begin{equation}
F_1(x) = \frac{1}{\eta}\exp\left(-\frac{2|x-\bar{x}|}{\eta}\right)\,\label{eq7}
\end{equation}
where $\bar x$ is the location parameter and $\eta > 0$ is the scale parameter (sometimes referred to as the diversity).

\subsection{Model 3: stochastic resetting}

The Brownian motion under Poissonian resetting can be described by a starightforward mathematical framework~\cite{em11,ross2014,mt22}. The Green function in the absence of resetting (or in other terms the propagator) is given by 
\begin{equation}
G_1(x,t|x_0)=\frac{1}{\sqrt{4\pi Dt}}\exp\left[-\frac{(x-x_0)^2}{4Dt}\right]\,,\label{eq8}
\end{equation}
where $x_0$ is the initial position, and $D$ is the diffusion constant. The initial condition takes the Dirac delta-function, namely $G_1(x,0|x_0)=\delta(x-x_0)$. Due to the Poissonian resetting with the rate $r$, the PDF $p_1(x,t|x_0)$ satisfies a renewal equation~\cite{ems20}. Its form is expressed by a sum of two terms, described as
\begin{eqnarray}
p_1(x,t|x_0)&=&e^{-rt}G_1(x,t|x_0)\nonumber\\
&+&r\int_0^te^{-r\tau}\,G_1(x,\tau|X_r)\,d\tau\,,\label{eq9}
\end{eqnarray}
where $X_r$  is the position to which the particle returns after resetting. It should be noticed that the renewal equation Eq.~(\ref{eq9}) holds for more general stochastic processes, with propagators having a more general form than $G_1(x,t|x_0)$~\cite{ems20,sw21}. In the stationary state with $t\to\infty$, the first term of Eq.~(\ref{eq9}) may be neglected, and the second term can be  exactly calculated~\cite{em11}, providing the typical Laplace PDF~\cite{kkp01}
\begin{equation}
p_1(x,\infty|x_0)=\frac{c_1}{2}e^{-c_1|x-X_r|}\,,\label{eq10}
\end{equation}
where the scale parameter $c_1=\sqrt{r/D}$ depends on the rate $r$. Hence, particles undergoing Brownian diffusion with resetting behave as performing confined motion with the Laplace statistics. Poissonian resetting also yields a similar behavior for subdiffusion~\cite{sw21}, i.e. random walks characterized by a nonlinear mean squared displacement $t^\alpha$ with $\alpha\in (0, 1)$. The only difference in this case, is that the scale parameter is given by $c_\alpha=\sqrt{r^\alpha/D}$. A similar result can also be obtained in a more general case. In fact, for a subordinator with an inverse infinitely-divisible distribution, the subordinated Brownian motion under Poissonian resetting tends to a stationary PDF in the Laplace form with the scale parameter $c_\Psi=\sqrt{\bar\Psi(r)/D}$, where the Laplace exponent $\bar\Psi(r)$ is expressed in terms of Bernstein functions~\cite{ssv10}. Therefore, Poissonian resetting can force many (but not any) nonstationary stochastic processes to a steady state with the Laplace PDF, and this general capability may explain why this distribution is often detected in confined trajectories of single particles. 

If, instead of Brownian motion, we consider the $\beta$-stable L\'evy motion with $\beta\in$ (0, 2) as a parent process~\cite{feller},  then  the Poissonian resetting of such random motion leads to a stationary characteristic function~\cite{sw21}
\begin{equation}
\hat{p}(k,\infty|X_r)=\frac{e^{ikX_r}}{1+D^*|k|^\beta/r}\,,\label{eq11}
\end{equation}
corresponding to the symmetric Linnik PDF~\cite{linnik63}. In Eq.~(\ref{eq11}) the term $e^{ikX_r}$ shows that the PDF maximum is located at $X_r$, and $D^*$ is constant. When $\beta=2$, the Linnik density reduces to the Laplace case. Both Laplace and Linnik PDFs describe confinement with jumps (unlike the Brownian confinement with continuous trajectories) \cite{sw21}, but the difference between them is that the Linnik PDF has a heavier tail than the Laplace one \cite{kkp01}. Therefore, the Linnik confinement is characterized by longer jumps.

\subsection{Advantages and disadvantages of models} 
As was shown above, there are different mathematical approaches leading to confinement. Although each model may be implemented in the dependence of physical conditions, they have their pros and cons, which are useful to list:
\begin{itemize}
\item subordination of random processes leads to the Laplace PDF in many cases of non-Gaussian processes, but the subordinator is too specific;
\item SDEs with an attractive potential are good for Gaussian confinement, but the Laplace PDF exists for the only value of the parameter describing the stationary PDF;
\item stochastic resetting has great potential for explaining non-equilibrium states in physics, chemistry and biology, and this model provides more grounds for understanding the diversity of confined trajectories.
\end{itemize}
The statistics of confined trajectories can play the key role in finding physical processes responsible for the occurrence of confinement. Therefore, discriminating tests of the random trajectories on possible PDFs are very important for the study of confined modes. It should be noticed that stochastic resetting is not so anomalous in living cells as it may seem. In particular, the diffusive backward motion of paused RNA polymerases is a diffusion process with stochastic resetting \cite{rlstg16}.

\begin{figure}
\centering
\includegraphics[width=\columnwidth]{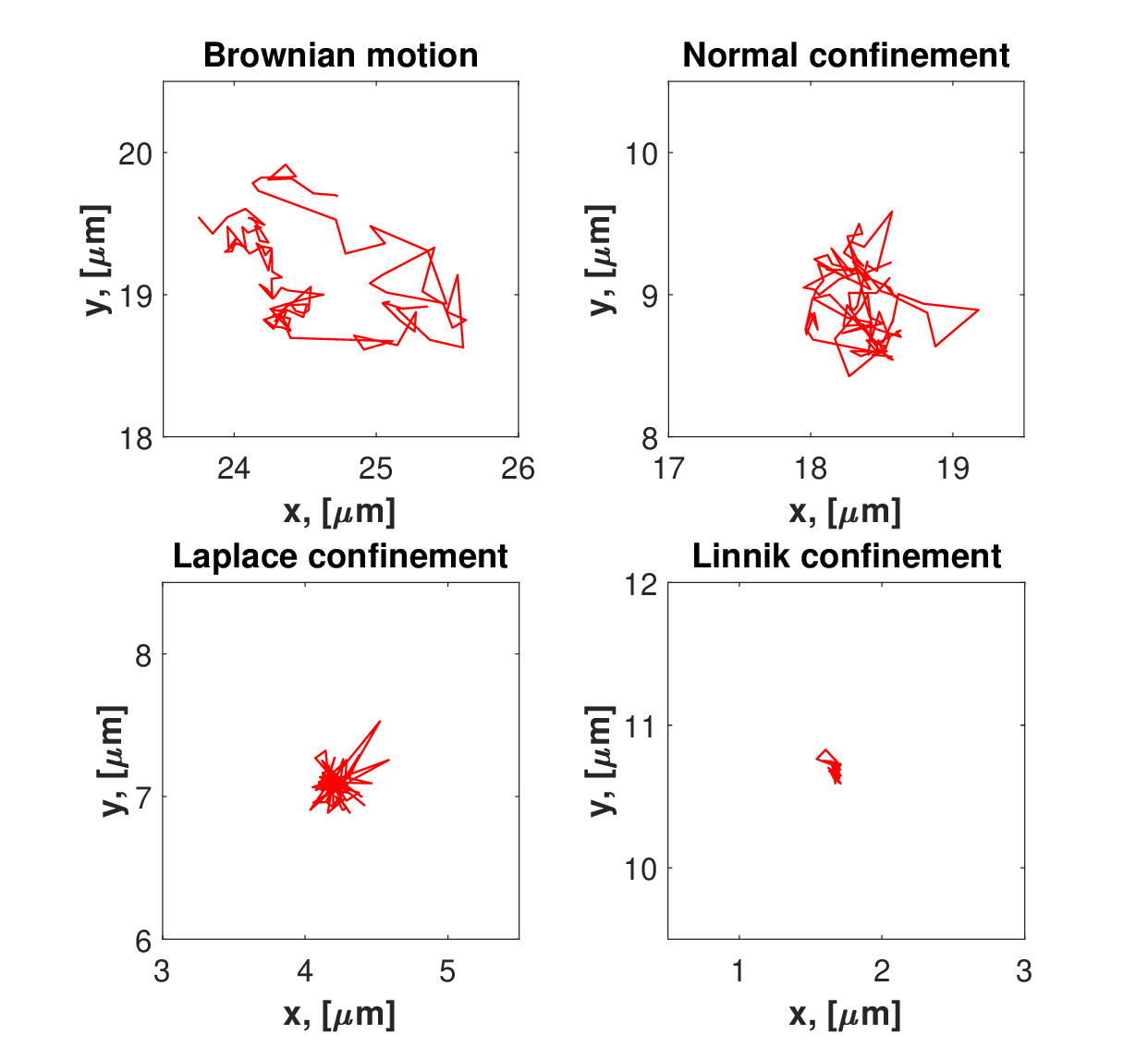}
\includegraphics[width=\columnwidth]{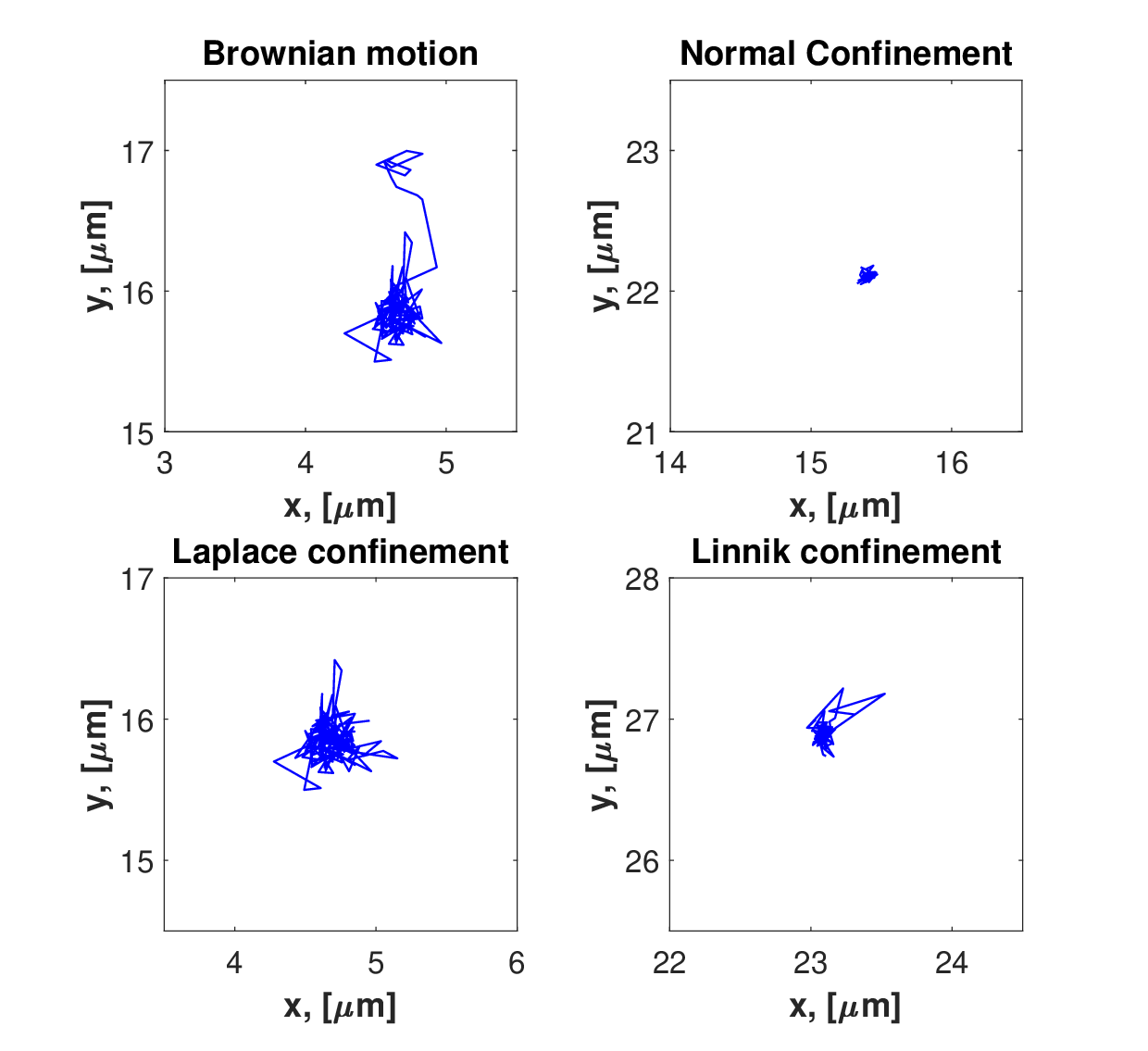}
\caption{\label{fig1} Examples of random one-segment trajectories, focusing on confined modes, in the motion of G proteins (a) and coupled receptors (b) observed in living cells. The type of segments is identified by the DC--MSS algorithm and statistical tests (see more details in Section~\ref{exp_data}).}
\end{figure}

\section{Analysis of experimental data}\label{exp_data}
To choose the most suitable model of confinement, we analyze stochastic trajectories detected in the SPT experiment \cite{sjbwlc17}.

\subsection{Classification of diffusion modes}\label{class_data} 

Truthful classification of random trajectories in cells is of great importance \cite{bswtk19}. At the current level of development of science and technology, it builds bridges between physics, biology, biochemistry, and biophysics necessary for understanding how living cells function on the microscopic basis \cite{mgp15}. We used the data of motion for G proteins and coupled receptors on the surface of living cells (Figure~\ref{fig1}). Their primary analysis can be started with classification of the trajectories according to the standardized maximal distance \cite{bkv18}. This method is useful because it allows us to evaluate quickly the contribution of confined trajectories \cite{sw20a,sw21}. However, it is rather rough and does not take into account the segmentation of trajectories. Segments can be different: immobile, confined, free (Brownian motion), and directed (diffusion with drift). Moreover, multi-segment trajectories show various modes in segments. Therefore, we used an accurate and computationally efficient transient motion analysis algorithm, termed ``divide-and-conquer moment scaling spectrum'' (DC--MSS) \cite{vfgj2018}. This approach includes three stages: initial track segmentation, initial segment classification, and final validation. The first stage calculates the maximum pairwise distance (MPD) between particle positions within the window. The value of MPD reveals the switches between different diffusive modes: MPD (immobile) $<$ MPD (confined) $<$ MPD (free) $<$ MPD (directed). In the next stage the track segments identified in the first stage are classified, using the MSS analysis of molecule displacements. The final stage compensates for initial track oversegmentation by testing the merger of adjacent segments. Using DC--MSS, we have carried out the segment analysis of trajectories for G proteins and coupled receptors in living cells. Their classification shows that there is a significant number of confined segments (see Table~\ref{table1}). The difference in the number of segments between G proteins and coupled receptors is due to a purely experimental situation: G proteins were presented with fewer trajectories than coupled receptors. Confined segments may have various statistics. Therefore, the next step of our data analysis is to discriminate them in statistics. 

\begin{table}
  \begin{center}
    \caption{Classification of experimental data with random-walk segments in the trajectories of G proteins and coupled receptors under basal conditions along the coordinates $x$ and $y$ with the cutoff length of trajectories more and equal to 50.}\label{table1}\vspace{1.5cm}
		\resizebox{0.48\textwidth}{!}{\begin{tabular}{|c|c|c||c|c|}
	\hline	
 \multicolumn{1}{|c}{\textbf{Types}}& \multicolumn{2}{|c||}{\textbf{G protein}} & \multicolumn{2}{c|}{\textbf{Coupled}}\\
\multicolumn{1}{|c}{\textbf{}}& \multicolumn{2}{|c||}{\textbf{}} & \multicolumn{2}{c|}{\textbf{Receptor}}\\
\hline
   \textbf{Coordinates} & \textbf{x} & \textbf{y} & \textbf{x} & \textbf{y} \\ 
\hline
    \textbf{Confined segm.} & 6513 & 6513 & 9637 & 9637\\
\hline
    \rowcolor{red!25}
  \textbf{Normal PDF} & 3373 & 3371 & 6684 & 6746\\
\hline
   \rowcolor{blue!25}
		\textbf{Laplace PDF} & 2245 & 2285 & 2353 & 2232\\
\hline
   \rowcolor{green!25}
   \textbf{Linnik PDF} & 895 & 857 & 600 & 659\\
\hline
\end{tabular}}
\end{center}
\end{table}

\subsection{Test of confined motion} 

To group confined tracks of G proteins and coupled receptors by statistics, we use a simple statistical test, based on the logarithm of the ratio of maximized likelihoods between normal and Laplace distributions \cite{kundu04}. The ratio $Q$ reads
\begin{equation}
Q = \frac{n}{2}\ln(2)-\frac{n}{2}\ln(\pi)+n\ln(\hat{\theta})-n\ln(\hat{\sigma})+\frac{n}{2}\,,\label{eq15}
\end{equation}
where the terms are dependent on the following values
\begin{eqnarray}
\hat{\theta}&=&\frac{1}{n}\sum_{i=1}^n |Y_i-\hat{\eta}|\,,\nonumber\\
\hat{\eta}&=&{\rm median}\{Y_1,Y_2,\dots,Y_n\}\,,\label{eq16}\\
\hat{\sigma}^2&=&\frac{1}{n}\sum_{i=1}^n (Y_i-\hat{\mu})^2\,,\quad
\hat{\mu}=\frac{1}{n}\sum_{i=1}^n Y_i\,.\nonumber
\end{eqnarray}
If the statistical test shows $Q>0$, the sample satisfies the normal distribution, but $Q<0$ suggests the Laplace distribution or a similar one to it. This procedure took into account only segments with a length greater than and equal to 50. Unfortunately, this test cannot refer segments with the Linnik statistics to the Laplace case due to their close relationship.

\subsection{Detection of Linnik confinement} 

To recognize segments with the Linnik PDF, we apply the test suggested by Anderson and Arnold \cite{aa93}. It allows estimating the parameter $\beta$ of the Linnik PDF for chosen segments. This test uses the minimization of the following objective function
\begin{equation}
\bar{I}_L=\int_0^\infty\left(\bar{\eta}(z)-(1+|\xi z|^\beta)^{-1}\right)^2\,\exp(-z^2)\,dz\,,\label{eq17}
\end{equation}
where $\bar{\eta}(z)=n^{-1}\sum_{j=1}^n\cos(z\,y_j)$, and $y_1, y_2, \dots, y_n$ is a data sampling. The expression is minimized with respect to two parameters, $\beta$ and $\xi$ (scale parameter). The presence of the weight function $\exp(-z^2)$ provides fast convergence of the integral. Thus, Eq. (\ref{eq17}) can be calculated numerically at no extra cost. The results of applying this test for G proteins and coupled receptors are shown in Table~\ref{table1}. Note that the diffusive motion of these particles differs in statistics along $x$ and $y$. Basically, the testing supports our above statistical analysis of confined segments. Really, the most segments with the jump-like confinement has the index $\beta$ equal or close to 2, typical for the Laplace statistics. Nevertheless, there is also a significant part of segments with $\beta\approx 1.203\pm 0.002$ corresponding to the Linnik case. They are detected in both cases, for G proteins and coupled receptors. This shows that the effect is not an anomaly or exotic.

\section{Resetting at power-law times}\label{mittag-leffler}
Let us mention that from the three models: subordination, system with potential and stochastic resetting only the last one can lead to the emergence of Linnik confinement from Brownian motion, represented certainly in many trajectories. Moreover, the stochastic resetting has the ability to be generalized from an exponential distribution of resetting times to a power-law one by using the Mittag-Leffler function. The Brownian motion with Poissonian resetting, leading to exponentially distributed  times, has been studied in great details (see \cite{em11,ems20} and references therein), while the power-law models have been explored less. For the first time stochastic resetting following the power-law distributed times, was considered in \cite{ng16,bcs19}. Below, we study such a model. It includes both the resetting with power-law times and the exponential case as a limit.

\begin{figure}
\centering
\includegraphics[width=\columnwidth]{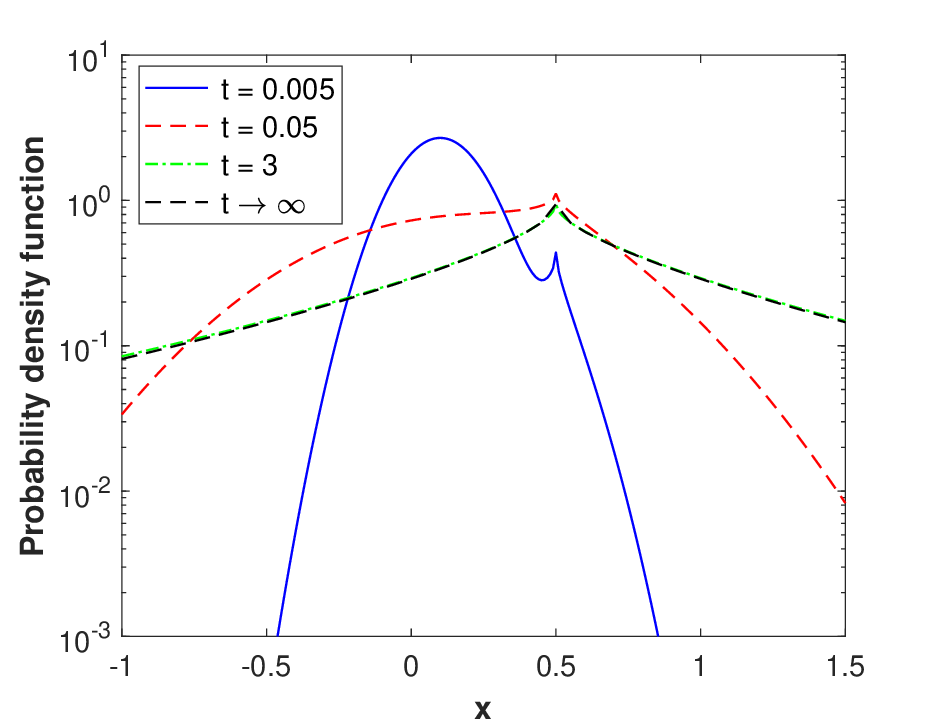}
\caption{\label{fig2} Brownian motion under the Mittag-Leffler resetting ($r = 3$, $D = 2$, $x_0 = 0.1$, $X_r = 0.5$, $\alpha=0.75$).}
\end{figure}

A more general resetting protocol is to take the sequence of resetting times to be generated by a probability density function $\psi(\tau)$. This is the so-called non-Poissonian resetting \cite{ems20}. It follows that $\Psi(\tau)=1-\int_0^\tau\psi(s)\,ds$ is the survival probability, i.e., the probability that no resetting has occurred up to time $\tau$. In particular, for Poissonian resetting, the functions read $\psi(\tau)=re^{-r\tau}$ and $\Psi(\tau)=e^{-r\tau}$. The first renewal equation (\ref{eq9}) can be easily generalized 
\begin{eqnarray}
p_\psi(x,t|x_0)&=&\Psi(t)\,G_1(x,t|x_0)\nonumber\\
&+&\int_0^t\psi(\tau)\,G_1(x,\tau|X_r)\,d\tau\,.\label{eq18}
\end{eqnarray}
The stochastic resetting with power-law times can be implemented with help of the Mittag-Leffler distribution. Let us recall that a statistical distribution in terms of the Mittag-Leffler function was defined by Pillai \cite{p90}. Differentiating it, the Mittag-Leffler PDF becomes $y^{\alpha-1}E_{\alpha,\alpha}(-y^\alpha)$ with $0<\alpha\leq 1$, where $E_{\alpha,\beta}(y)=\sum_{k=0}^\infty y^k/\Gamma(\alpha k+\beta)$ is the two-parameter Mittag-Leffler function~\cite{mh08}. If $\alpha\to 1$, the PDF tends to exponential. Next, in a stationary state of Eq. (\ref{eq18}) the PDF $p_\alpha(x,t|x_0)$ takes the following form
\begin{eqnarray}
p_\alpha(x,\infty|x_0)&=&r\int_0^\infty (r\tau)^{\alpha-1}E_{\alpha,\alpha}[-(r\tau)^\alpha]\nonumber\\
&\times&G_1(x,\tau|X_r)\,d\tau\,.\label{eq19}
\end{eqnarray}
Note that the normalization condition $\int_{-\infty}^\infty p_\alpha(x,\infty|x_0)\,dx =1$ is connected with the integral $\int_0^\infty (r\tau)^{\alpha-1}E_{\alpha,\alpha}[-(r\tau)^\alpha]\,d\tau = 1/r$. Moreover, in this context the function $\Psi(t)=E_{\alpha,1}[-(rt)^\alpha]$ tends to $e^{-rt}$ for $\alpha\to 1$ and, as expected, the relation $r=0$ shows the absence of resetting in Eq.~(\ref{eq18}). If we pass from the PDF $G_1(x,t|x_0)$ to its characteristic function $e^{-k^2Dt}$ through the Fourier transform, then it is not difficult to find the characteristic function for $p_\alpha(x,\infty|x_0)$, namely
\begin{eqnarray}
\hat p_\alpha(k,\infty|x_0)&=&\int_{-\infty}^\infty p_\alpha(x,\infty|x_0)\,e^{ikx}\,dx\nonumber\\
&=&\frac{e^{ikX_r}}{1+D^\alpha k^{2\alpha}/r^\alpha}\,.\label{eq20}
\end{eqnarray}
When $\alpha=1$, the expression $\hat p_1(k,\infty|x_0)$ is the characteristic function for Eq. (\ref{eq10}), clearly describing the resetting with exponentially distributed times. After the inverse Fourier transform, the characteristic function (\ref{eq20}) leads to the Linnik distribution generalizing the Laplace distribution \cite{kkp01}. The evolution from the Brownian motion under non-Poissonian resetting to the Linnik distribution is presented in Fig.~\ref{fig2}. Thus, the Linnik distribution as a stationary state in random processes under stochastic resetting can occur not only, when the $\beta$-stable L\'evy motion for $0<\beta\leq2$ is subject to the Poissonian resetting \cite{sw21,bchpm23}, but also when Brownian motion is under resetting at power-law times generated by the Mittag-Leffler distribution in which $0<\alpha\leq1$.

The model of the Mittag-Leffler resetting is characterized by the same index for small and large times. In Appendix~\ref{appA} we present its generalization, using the three-parameter Mittag-Leffler function, where the stochastic resetting depends on two indices. This makes the model more flexible and more adequate for the stochastic resetting behaving differently for small and large times. Moreover, Appendix~\ref{appB} considers the optimal search for the Mittag-Leffler resetting.

\section{Conclusions} 
Single molecular imaging of G proteins and coupled receptors \cite{cklmb21} as they diffuse and interact on the surface of a living cell shows many diffusion modes with frequent switching between them. Confined segments of their trajectories are typical for the interaction of individual coupled receptors and G proteins. Their behavior resembles the stochastic resetting. The coupled receptors have to ``find'' and interact with G proteins on the cell membrane in order to initiate and regulate intracellular processes \cite{tlygdf12}. The processes are highly heterogeneous and complex. The interaction between G proteins and coupled receptors occurs multiple times, and the time interval between activation and deactivation turns out to be random. The stochastic resetting makes this process optimal. If the stochastic resetting happens with the exponentially distributed times, then the confined motion obeys the Laplace distribution, but power-law distributed times generate the Linnik distribution in confined modes. We analyzed the segmented experimental data with random trajectories of coupled receptors and G proteins. Among the set of segments, by using special tests, we found confined ones with the Laplace and Linnik statistics. Our analysis to confined trajectories and fragments can serve as a filter to identify the interaction of G proteins and coupled receptors at power-law times. 

Finally, we can summarize the following conclusions.
\begin{itemize}
\item[(i)] The binding of G proteins and coupled receptors \cite{wk18} causes their confinement.
\item[(ii)] The resetting corresponds to rebinding events. If this is the case, it means that the confined segments actually correspond to binding-unbinding-rebinding events where the unbinding step has not been detected.
\item[(iii)] If the time spent unbound is short (the DC-MSS cannot detect segments shorter than 20 steps), then assuming that between two consecutive binding events the G proteins undergo Brownian motion, the Linnik/Laplace statistics might just be the result of the mixing due to the suboptimal segmentation or the presence of unbinding events shorter than the minimum detectable duration.
\end{itemize}

\begin{acknowledgments} 
A.W. thanks the support of Beethoven Grant No.  DFG-NCN 2016/23/G/ST1/04083. The authors are grateful to T. Sungkaworn and D. Calebiro for free access to their experimental data analyzed in Section~\ref{exp_data} as well as to C. Manzo for useful discussions.
\end{acknowledgments}

\appendix
\section{Generalized Mittag-Leffler resetting}
\label{appA}
The three-parameter Mittag–Leffler function $E_{\alpha,\beta}^\gamma(y) $ is a generalization of the two-parameter Mittag–Leffler function $E_{\alpha,\beta}(y)$ mentioned in Section~\ref{mittag-leffler}. It was first introduced by Prabhakar \cite{prab71} who defined it as
\begin{equation}
E_{\alpha,\beta}^\gamma(y) = \sum_{k=0}^\infty \frac{(\gamma)_k\,y^k}{k!\Gamma(\alpha k+\beta)}\,,\label{eq1A}
\end{equation}
where $k\in\mathbb{N}$, and $(\gamma)_k:= \Gamma(k+\gamma)/\Gamma(\gamma)$ is the Pochhammer symbol. A random variable $Y$ has a generalized Mittag–Leffler distribution \cite{jwt08,juslh10,cp13}, if its probability density function is
\begin{equation}
p_Y(y) = \lambda^\delta y^{\delta\nu-1}E_{\nu,\delta\nu}^\delta(-\lambda y^\nu)\label{eq2A}
\end{equation}
with $y>0$, $\nu\in(0,1]$, shape $\delta\in\mathbb{R}$ and rate $\lambda>0$. Then the first renewal equation (\ref{eq18}) reads
\begin{eqnarray}
&&p_{\alpha,\delta}(x,t|x_0)\nonumber\\
&=&\left(1-(r\tau)^{\delta\alpha}E_{\alpha,\delta\alpha+1}^\delta[-(r\tau)^\alpha]\right)G_1(x,t|x_0)\nonumber\\
&+&r\int_0^t(r\tau)^{\delta\alpha-1}E_{\alpha,\delta\alpha}^\delta[-(r\tau)^\alpha]\,G_1(x,\tau|X_r)\,d\tau .\label{eq3A}
\end{eqnarray}
In a stationary state of Eq. (\ref{eq3A}), the PDF $p_{\alpha,\delta}(x,t|x_0)$ is easier to find through its characteristic function, namely
\begin{eqnarray}
\hat p_{\alpha,\delta}(k,\infty|x_0)&=&\int_{-\infty}^\infty p_{\alpha,\delta}(x,\infty|x_0)\,e^{ikx}\,dx\nonumber\\
&=& \frac{e^{ikX_r}}{\left(1+D^\alpha k^{2\alpha}/r^\alpha\right)^\delta}\,.\label{eq4A}
\end{eqnarray}
As a result, we obtain the characteristic function of the generalized Linnik PDF \cite{kkp01}. When $\delta=1$, the PDF $p_{\alpha,\delta}(x,\infty|x_0)$ simplifies to the ordinary Linnik form (see Section~\ref{mittag-leffler}). If $\alpha=1$, then the stationary PDF $p_{1,\delta}(x,\infty|x_0)$ can be found explicitly in the form (\ref{eq5}) that is the same for the coupled Langevin equations mentioned in Subsection~\ref{sdes}. 

\section{Mean time to absorption}
\label{appB}
For the non-Poissonian resetting there is a convenient way to connect the survival probability with resetting, denoted as $Q_r$, while one without resetting as $Q_1$ \cite{ems20}. It is based on the last renewal equation written as
\begin{eqnarray}
Q_r(x_0,t)&=&\Psi(t)Q_1(x_0,t)\nonumber\\
&+&\int_0^t\psi(\tau)\,Q_1(X_r,\tau)\,Q_r(x_0,t-\tau)\,d\tau\label{eq1B}
\end{eqnarray}
in the shorthand $Q_r(x_0, t|X_r) = Q_r(x_0,t)$ and similarly for $Q_1$. Taking the Laplace transform in time and setting $x_0 = X_r$, we have
\begin{equation}
\bar{Q}_r(X_r,s)=\frac{\int_0^\infty e^{-st}\,\Psi(t)\,Q_1(X_r,t)\,dt}{1-\int_0^\infty e^{-st}\,\psi(t)\,Q_1(X_r,t)\,dt}\,.\label{eq2B}
\end{equation}
Then the mean time to absorption is obtained by setting $s = 0$ in Eq. (\ref{eq2B}), i.\,e.
\begin{equation}
\langle T(X_r)\rangle=\frac{\int_0^\infty \Psi(t)\,Q_1(X_r,t)\,dt}{1-\int_0^\infty \psi(t)\,Q_1(X_r,t)\,dt}\,.\label{eq3B}
\end{equation}
The survival probability of a diffusive particle starting from $x_0$ without resetting is written as 
\begin{equation}
Q_1(x_0,t)=\mathrm{erf}\left(\frac{x_0}{2(Dt)^{1/2}}\right)\,,\label{eq4B}
\end{equation}
where $\mathrm{erf}(z)=\frac{2}{\sqrt{\pi}}\int_0^ze^{-y^2}\,dz$ is the Gauss error function \cite{as65}. In Eq. (\ref{eq3B}) only the functions $\Psi$ and $\psi$ depend on $r$. Therefore, by changing variables $rt\to t$ it is convenient to collect all parameters ($r$, $X_r$, and $D$) in $Q_1(X_r,t)$, i.e.,
\begin{equation}
Q_1(X_r,t)=\mathrm{erf}\left(\frac{\gamma}{2t^{1/2}}\right)\,,\label{eq5B}
\end{equation}
where $\gamma=X_r\sqrt{r/D}$, taking $\Psi(t,r=1)$ and $\psi(t,r=1)$ for the fixed $r$. Consequently, Eq. (\ref{eq3B}) yields
\begin{equation}
\langle T(X_r)\rangle=\frac{1}{r}\left(\frac{\int_0^\infty \Psi(t,r=1)\,Q_1(X_r,t)\,dt}{1-\int_0^\infty \psi(t,r=1)\,Q_1(X_r,t)\,dt}\right)\,.\label{eq6B}
\end{equation}
Since $\mathrm{erf}(z)\sim 2z/\sqrt{\pi}$ for $z$ tending to zero, the value of $\langle T(X_r)\rangle$ diverges as $r\to 0$ as $\langle T(X_r)\rangle\sim r^{-1/2}$. It is clear that in the absence of resetting, the mean time is infinity. On the other hand, when $r\to\infty$, the survival probability without resetting $Q_1(X_r,t)$ is limited to one. Then $\int_0^\infty \psi(t,r=1)\,dt\to 1$, and the asymptotic behavior of the denominator in Eq. (\ref{eq6B}) is determined by $1-\mathrm{erf}(z)\sim e^{-z^2}/(\sqrt{\pi}z)$ for $z\to\infty$. Therefore, $\langle T(X_r)\rangle$ also diverges as $r\to\infty$. This is not surprising. The higher the reset rate, the shorter the time between resets to reach the origin. These two divergences can surround one minimum of $\langle T(X_r)\rangle$. 

\begin{figure}
\centering
\includegraphics[width=\columnwidth]{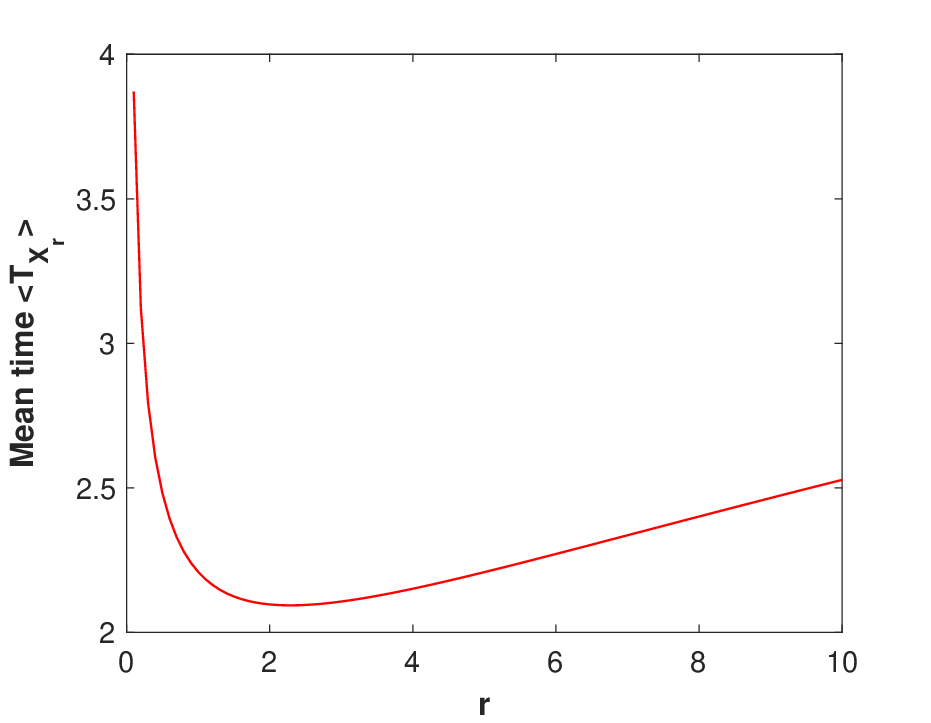}
\caption{\label{fig4} Mean time to absorption as a function of $r$  with the generalized Mittag-Leffler resetting ($D = 1$, $x_0 = X_r = 1$, $\alpha=0.85$ and $\delta=0.5$).}
\end{figure}

However, this will be the case if the numerator of Eq. (\ref{eq6B}) is finite. Note that $\int_0^\infty \psi(t,r=1)\,Q_1(X_r,t)\,dt<1$ for $r<\infty$. Taking $\Psi(t,r=1)=E_{\alpha,1}(-t^\alpha)$, the improper integral in the numerator of Eq. (\ref{eq6B}) converges, when $\alpha>1/2$. Here the convergence problem arises at the upper limit of the integral. For $t\to\infty$ the integrand $\Psi(t,r=1)\,Q_1(X_r,t)$ depends on $E_{\alpha,1}(-t^\alpha)\sim t^{-\alpha}/\Gamma(1-\alpha)$, having $0<\alpha\leq 1$, and $Q_1(X_r,t)\sim\gamma/\sqrt{\pi t}$. This leads to the convergence of the numerator of Eq. (\ref{eq6B}) just as in the case of $1/2<\alpha\leq 1$. When $\Psi(t,r=1)$ corresponds to the generalized Mittag-Leffler resetting, we observe the same convergence condition of this numerator, i.\,e. for $1/2<\alpha\leq 1$. If the numerator of Eq. (\ref{eq6B}) is finite, between two divergences ($r\to 0$ and $r\to\infty$) there is a single minimum. Such an example (related to the generalized Mittag-Leffler resetting) is shown in Figure~\ref{fig4}. In this case the minimal mean time is achieved when the ratio of the distance $X_r$ (from the resetting position to the target) to the typical diffusion length between resets is $\gamma=1.5124\dots$. Finally, the Mittag-Leffler (generalized and ordinary) resetting with $0<\alpha\leq 1/2$ is characterized by divergence, and the mean time for a diffusive particle to reach the origin becomes infinite for any $r$.


\begin{thebibliography}{2021}

\bibitem{mgp15}
C. Manzo and M.F. Garcia-Parajo, A review of progress in single particle tracking, Rep. Prog. Phys. {\bf 78}, 124601 (2015).

\bibitem{sjbwlc17}
T. Sungkaworn, M-L. Jobin, K. Burnecki, A. Weron, M. J. Lohse, and D. Calebiro, Single-molecule imaging reveals receptor-G protein interactions at cell surface hot spots, Nature (London) {\bf 550}, 543 (2017).

\bibitem{lgkc21}
Y. Lanoisel\'ee, J. Grimes, Z. Koszegi, and D. Calebiro, Detecting transient trapping from a single trajectory: A structural approach, Entropy {\bf 23}, 1044 (2021).

\bibitem{mgvgm21}
G. Mu$\tilde{\rm n}$oz-Gil, G. Volpe, M.A. Garcia-March, et al., Objective comparison of methods to decode anomalous diffusion, Nat. Commun. {\bf 12}, 6253 (2021).

\bibitem{csc14}
N. Chenouard, I. Smal, F. de Chaumont, et al., Objective comparison of particle tracking methods, Nat. Methods {\bf 11}, 281 (2014).

\bibitem{sj97}
M.J. Saxton and K. Jacobson, Single-particle tracking: applications to membrane dynamics, Annu. Rev. Biophys. Biomol. Struct. {\bf 26}, 373 (1997).

\bibitem{w09}
D. Wirtz, Particle-tracking microrheology of living cells: principles and applications, Annu. Rev. Biophys. {\bf 38}, 301 (2009).

\bibitem{bz07}
B. Brandenburg and X. Zhuang, Virus trafficking -- learning from single-virus tracking, Nat. Rev. Microbiol. {\bf 5}, 197 (2007).

\bibitem{msm97}
W.F. Marshall, A. Straight, J.F. Marko, et al., Interphase chromosomes undergo constrained diffusional motion in living cells, Curr. Biol. {\bf 7}, 930 (1997).

\bibitem{bkv18}
V. Briane, C. Kervrann and M. Vimond, Statistical analysis of particle trajectories in living cells, Phys. Rev. E {\bf 97}, 062121 (2018).

\bibitem{wjbsc18}
A. Weron, J. Janczura, E. Boryczka, T. Sungkaworn, and D. Calebiro, Statistical testing approach for fractional anomalous diffusion classification, Phys. Rev. E 99, 042149 (2019).

\bibitem{jklosw20}
J. Janczura, P. Kowalek, H. Loch-Olszewska, J. Szwabi\'nski, A. Weron,  Classification of particle trajectories in living cells: Machine learning versus statistical testing hypothesis for fractional anomalous diffusion, Phys. Rev. E 102, 032402 (2020).

\bibitem{bkjbgw12}
K. Burnecki, E. Kepten, J. Janczura, I. Bronshtein, Y. Garini, and A. Weron, Universal algorithm for identification of fractional Brownian motion. A case of telomere subdiffusion, Biophys. J. {\bf 103}, 1839
(2012).

\bibitem{sw20a} 
A. Stanislavsky and A. Weron, Look at tempered subdiffusion in a conjugate map: Desire for the confinement, Entropy {\bf 22}, 1317 (2020).

\bibitem{ewowals12}
F.A. Espinoza, M.J. Wester, J.M. Oliver, B.S. Wilson, N.L. Andrews, D.S. Lidke, and S.L. Steinberg, Insights into cell membrane microdomain organization from live cell single particle tracking of the IgE high affinity receptor Fc$\epsilon$RI of mast cells, Bull. Math. Biol. {\bf 74}, 1857 (2012).

\bibitem{sw21a}
A. Stanislavsky and A. Weron, Confined random motion with Laplace and Linnik statistics, J. Phys. A: Math. Theor. {\bf 54}, 055009 (2021).

\bibitem{ou30}
G.E. Uhlenbeck and L.S. Ornstein, On the theory of Brownian motion, Phys. Rev. {\bf 36} (5), 823 (1930).

\bibitem{em11}
M.R. Evans and S.N. Majumdar, Diffusion with stochastic resetting, Phys. Rev. Lett. {\bf 106}, 160601 (2011).

\bibitem{ross2014}
S.M. Ross, {\it Introduction to Probability Models}, 11th Edition (Academic Press, New York, 2014).

\bibitem{mt22}
M. Magdziarz and K. Ta\'zbirski, Stochastic representation of processes with resetting, Phys. Rev. E {\bf 106}, 014147 (2022).

\bibitem{ems20}
M.R. Evans, S.N. Majumdar, G. Schehr, Stochastic resetting and applications, J. Phys. A: Math. Theor. {\bf 53}, 193001 (2020).

\bibitem{pkr20}
A. Pal, L. Ku\'smierz, and S. Reuveni, Search with home returns provides advantage under high uncertainty, Phys. Rev. Research {\bf 2}, 043174 (2020).

\bibitem{sw22}
A. Stanislavsky and A. Weron, Subdiffusive search with home returns via stochastic resetting: A subordination scheme approach, J. Phys. A: Math. Theor. {\bf 55}, 074004 (2022).

\bibitem{sw21}
A. Stanislavsky and A. Weron, Optimal non-Gaussian search with stochastic resetting, Phys. Rev. E {\bf 104}, 014125 (2021).

\bibitem{kkp01}
S. Kotz, T. Kozubowski, and K. Podg\'orski, {\it The Laplace Distribution and Generalizations: A Revisit with Applications to Communications, Economics, Engineering, and Finance} (Birkhauser, Boston, 2001).

\bibitem{b49}
S. Bochner, Diffusion equation and stochastic processes, Proc. Natl. Acad. Sci. U.S.A. {\bf 35}, 368 (1949).

\bibitem{sw20}
A. Stanislavsky and A. Weron, Accelerating and retarding anomalous diffusion: A Bernstein function approach, Phys. Rev. E {\bf 101}, 052119 (2020).

\bibitem{lscw22}
Y. Lanoisel\'ee, A. Stanislavsky, D. Calebiro, and A. Weron, Temperature and friction fluctuations inside a harmonic potential, Phys. Rev. E {\bf 106}, 064127 (2022).

\bibitem{cir85}
J.C. Cox, J.E. Ingersoll, and S.A. Ross, A theory of the term structure of interest rates, Econometrica {\bf 53}(2), 385 (1985).

\bibitem{brm2020}
S.P. Blomberg, S.I. Rathnayake, and C.M. Moreau, Beyond Brownian motion and the Ornstein-Uhlenbeck process: stochastic diffusion models for the evolution of quantitative characters, The American Naturalist {\bf 195}(2), 145-165 (2020).

\bibitem{as65}
M. Abramowitz and I.A. Stegun, {\it Handbook of Mathematical Functions: With Formulas, Graphs, and Mathematical Tables} (Dover Publications, New York, 1965).

\bibitem{k82}
O. Krop\'a\v c, Some properties and applications of probability distributions based on MacDonald
function, Aplikace matematiky {\bf 27}(4), 285–302 (1982).


\bibitem{ssv10}
R. L. Schilling, R. Song, and Z. Vondra$\check{\rm c}$ek, {\it Bernstein Functions: Theory and Applications} (de Gruyter Studies, Berlin, 2010).

\bibitem{feller}
W. Feller, {\it Introduction to Probability Theory and Its Application}, Vol. II (John Wiley \& Sons, New York, 1967).

\bibitem{linnik63}
Yu.V. Linnik, Linear forms and statistical tests, Ukrain. Math. Z. {\bf 5}, pp. 207-243, {\it English Translations in Mathematical statistics and Probability}, American Mathematical Society, Providence, R.I., {\bf 3}, pp. 1-40, pp. 41-90 (1962).

\bibitem{rlstg16}
\'E. Rold\'an, A. Lisica, D. S\'anchez-Taltavull, and S. W. Grill, Stochastic resetting in backtrack recovery by RNA polymerases, Phys. Rev. E {\bf 93}, 062411 (2016).

\bibitem{bswtk19}
K. Burnecki, G. Sikora, A. Weron, M.M. Tamkun, and D. Krapf, Identifying diffusive motions in single-particle trajectories on the plasma membrane via fractional time-series models, Phys. Rev E {\bf 99}, 012101 (2019).

\bibitem{vfgj2018}
A.R. Vega, S.A. Freeman, S. Grinstein, and Kh. Jaqaman, Multistep track segmentation and motion
classification for transient mobility analysis, Biophys. J. {\bf 114}, 1018 (2018).

\bibitem{kundu04}
D. Kundu, Discriminating between normal and Laplace distributions, In: N. Balakrishnan, H.N. Nagaraja and N. Kannan (eds) {\it Advances in Ranking and Selection, Multiple Comparisons, and Reliability. Statistics for Industry and Technology} (Boston, Birkh\"auser, 2005) pp. 65--79.

\bibitem{aa93}
D.N. Anderson and B.C. Arnold, Linnik distributions and processes, J. Appl. Probab. {\bf 30}(2), 330 (1993).

\bibitem{ng16}
A. Nagar and S. Gupta, Diffusion with stochastic resetting at power-law times, Phys. Rev. E {\bf 93}, 060102(R) (2016).

\bibitem{bcs19}
A.S. Bodrova, A.V. Chechkin, and I.M. Sokolov, Nonrenewal resetting of scaled Brownian motion, Phys. Rev. E {\bf 100}, 012119 (2019);
Scaled Brownian motion with renewal resetting, Phys. Rev. E {\bf 100}, 012120 (2019).

\bibitem{p90}
R.N. Pillai, On Mittag-Leffler functions and related distributions, Ann. Inst. Stat. Math. {\bf 42}, 157 (1990).

\bibitem{mh08}
A.M. Mathai and H.J. Haubold, {\it Special Functions for Applied Scientists} (Springer, Berlin, 2008).

\bibitem{bchpm23}
C. Di Bello, A.V. Chechkin, A.K. Hartmann, Z. Palmowski, R. Metzler, Time-dependent probability density function for partial resetting 
dynamics, New J. Phys. {\bf 25}, 082002 (2023). 

\bibitem{cklmb21}
D. Calebiro, Z. Koszegi, Y. Lanoisel\'ee, T. Miljus, and S. O'Brien, G protein-coupled receptor-G protein interactions: a single-molecule perspective, Physiol. Rev. {\bf 101}(3), 857 (2021). 

\bibitem{tlygdf12}
B. Trzaskowski, D. Latek, S. Yuan, U. Ghoshdastider, A. Debinski, and S. Filipek, Action of Molecular Switches in GPCRs -- Theoretical and Experimental Studies, Curr. Med. Chem. {\bf 19}(8), 1090 (2012).

\bibitem{wk18}
W.I. Weis and B.K. Kobilka, The molecular basis of G protein–coupled receptor activation, Annu. Rev. Biochem. {\bf 87}, 897 (2018).

\bibitem{prab71}
T.R. Prabhakar, A singular integral equation with a generalized Mittag–Leffler function in the kernel, Yokohama Math. J. {\bf 19}, 7 (1971).

\bibitem{jwt08}
A. Jurlewicz, K. Weron, and M. Teuerle, Generalized Mittag-Leffler relaxation: Clastering-jump continous-time random walk approach, Phys. Rev. E {\bf 78}, 011103 (2008).

\bibitem{juslh10}
K.K. Jose, P. Uma, V. Seetha Lekshmi, and H.J. Haubold, Generalized Mittag-Leffler distributions and processes for applications in astrophysics and time series modeling, In: {\it Proceedings of the Third UN/ESA/NASA Workshop on the International Heliophysical Year 2007 and Basic Space Science}, Eds. H.J. Haubold and A.M. Mathai, (Springer, Berlin Heidelberg, 2010), pp. 79-92.

\bibitem{cp13}
D.O. Cahoy and F. Polito, Renewal processes based on generalized Mittag–Leffler waiting times, Commun. Nonlinear Sci. Numer. Simul. {\bf 18}(3), 639 (2013).


\end{thebibliography}
\end{document}